# All-fiber dual-wavelength mode-locked laser using a bend-induced-birefringence Lyot-filter as gain-tilt equalizer

Yuanjun Zhu, Fulin Xiang, Lei Jin, Sze Yun Set*, Shinji Yamashita

Research Center for Advanced Science and Technology, The University of Tokyo, 4-6-1, Komaba, Meguro-ku, Tokyo 153-8904, Japan

Corresponding Author: set@cntp.t.u-tokyo.ac.jp

**Abstract:** Multi-wavelength fiber lasers have emerged as a promising light source for the application in wavelength division multiplexing communication, terahertz wave generation and optical sensing due to high efficiency and robustness. Lyot-filter emerges as a potential device for multi-wavelength generation application. However, because of the high birefringence of polarization maintaining fiber in Lyot-filter, it is difficult to generate broadband dual-wavelength or multi-wavelength mode-locked fiber laser by using common Lyot-filter. In this paper, for the first time, we propose an idea of using the low birefringence induced by bending the single-mode fiber to form a Lyot-filter for dual-wavelength mode-locked fiber laser generation. The dual-wavelength output centers at 1532 and 1556 nm and it may simplify the set-up for dual-wavelength mode-locked laser.

**Index Terms:** Mode-locked laser, Fiber lasers.

## 1. Introduction

Dual-wavelength or multi-wavelength fiber lasers have the advantages of high efficiency, compact structure and robustness so that they have found applications in optical fiber sensing, laser comb generation, wavelength-division-multiplexed communication and terahertz wave generation [1-4]. Comb filters [5] such as Lyot-filters [6], Sagnac loop filters [7], fiber Bragg gratings (FBGs) [8], Fabry-Perot filters [9] and Mach-Zehnder interferometer [10] are commonly used devices for multi-wavelength generation. However, in most cases, the free spectrum range (FSR) of these devices are relatively narrow, which makes them unsuitable for mode-locked lasers which typically operates with a broad output spectrum. In order to generate multi-wavelength outputs in a mode-locking laser, many researchers combined a comb filter together with mode-locking techniques such as nonlinear polarization rotator (NPR), nonlinear optical loop mirror (NOLM) and nonlinear amplification loop mirror (NALM) [11-17]. Because of gain competition between each of these multi-wavelength lasing modes, together with the delicate condition of the mode-locking techniques, it is not easy to satisfy these conditions for stable operation.

The erbium-doped fiber (EDF) has two inhomogeneously broadened gain peaks, typically at 1532nm and 1556nm. However, because of the gain competition between these two peaks, only a single-wavelength mode-locked output can be achieved. One technique was proposed using a programmable attenuator to effectively control the saturation condition and the gain profile of the EDF in the laser cavity through intra-cavity loss tuning [18]. However, the gain profile of an EDF is quite complex and difficult to be controlled by setting just a constant loss parameter with a programmable attenuator. Therefore high quality stable dual-wavelength mode-locking is hard to be achieved.

Among these techniques for dual-wavelength or multi-wavelength mode-locking, Lyot-filter (birefringence-induced spectral filter) is a common approach because of its compact structure and broadband operation property [19-21]. A typical Lyot-filter which uses polarization maintaining fiber (PMF) as birefringent fiber has a narrow FSR which is usually as narrow as several nm. Here we proposed a broad FSR, low-birefringent Lyot-filter realized by simply bending a length of standard single-mode fiber (SMF) to provide the birefringence required. The proposed Lyot-filter has a much wider FSR and pass-band, which makes it more suitable for dual-wavelength mode-locked generation. In this paper, we demonstrate a dual-wavelength mode-locked laser at 1532 nm and 1556 nm, with pulse widths of 940 fs and 820 fs, respectively, using a bend-induced-birefringence Lyot-filter. We believe it can potentially to be used in dual-comb spectroscopy and terahertz wave generation. By appropriately adjusting the polarization controllers (PCs), single-wavelength mode-locked outputs center at either 1532 nm or 1556 nm can also be selectively tuned.

## 2. Experimental Details

A commonly used fiber-based Lyot-filter is composed of a section of PMF, a polarizer and a PC. The transmission property can be calculated using Jones matrix as follows [22]:

$$\boldsymbol{E}_{out} = \boldsymbol{M} * \boldsymbol{P} * \boldsymbol{J} * \boldsymbol{E}_{in} \tag{1}$$

where $E_{out}$ and $E_{in}$ are the output and input signal. $M$, $P$ and $J$ are the Jones matrices of the polarizer, the PC and the PMF respectively. The transmission function of the Lyot-filter can be written as:

$$T = \frac{|E_{out}|^2}{|E_{in}|^2} = \frac{1}{2}cos^2(\frac{\pi\Delta n}{\lambda}L_{PMF})(1+sin2\theta) \qquad (2)$$

where the Δn is the refractive index difference between the fast axis and slow axis of the PMF, θ is the angle between the polarization direction of the input light and the fast axis of the PMF, which can be tuned by adjusting the PC, $L_{PMF}$ is the length of the PMF. The FSR of the transmission function can be expressed as:

$$FSR \approx \frac{\lambda^2}{\Delta n L_{PMF}} \qquad (3)$$

From the equation (3), the FSR of Lyot-filter is inversely proportional to the birefringence and the length of the PMF. The birefringence of a length of PMF is usually rather large, therefore giving a narrow FSR, which makes it unsuitable for broadband multi-wavelength or dual-wavelength mode-locking of a fiber laser.

One could design a Lyot-filter with a broad FSR using either a short length of PMF or a length of low-birefringence fiber. Compared with using a short length PMF, using the low-birefringence by bending the SMF [23-25] to form a bend-induced birefringence Lyot-filter is simpler and easier to be adjusted. Because it is not easy to precisely control the length of PMF in Lyot-filter. The bend-induced-birefringence is composed of a PC, a 25-m-long SMF which is bent with the diameter of 13 cm and a polarizer. The configuration of the test system of bend-induced-birefringence transmission property is illustrated in Fig. 1. A commercial erbium-doped fiber amplifier (EDFA) acts as the input source which passes through a polarizer to be linearly polarized and then split into two equal intensity sources via a 3 dB coupler. Port 1 of the 3dB coupler is connected directly to an optical spectrum analyzer (OSA) (YOKOGAWA AQ6375) as a reference signal while port 2 connected to the bend-induced birefringence and then to the OSA. The spectral transmission characteristics are measured by dividing the output spectrum with the reference spectrum.

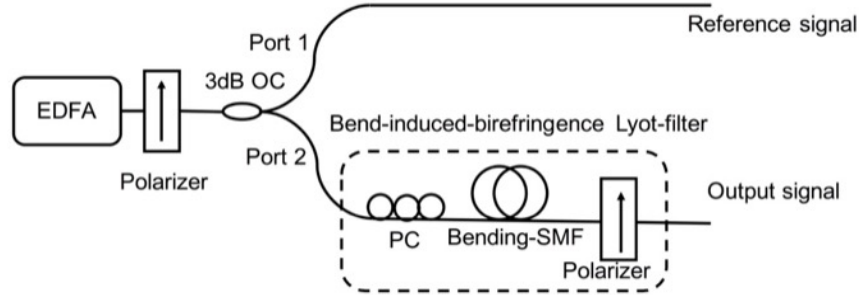

Fig. 1. (a) Diagram of the experimental set up. (b) Detail of the test object used

The detected transmission result is shown in the Fig. 2, showing the transmission profiles of the six different states which correspond to different positions of the PC. The traces show different slope and amplitude, which indicates that this device can introduce gain-tilt to equalize the gain competition between the inhomogeneously broaden gain peaks of an EDF. Therefore, by adjusting the PC in the laser cavity, it is possible to balance the gain spectrum of the EDF to achieve simultaneous lasing at the two gain peaks at 1532 nm and 1556 nm.

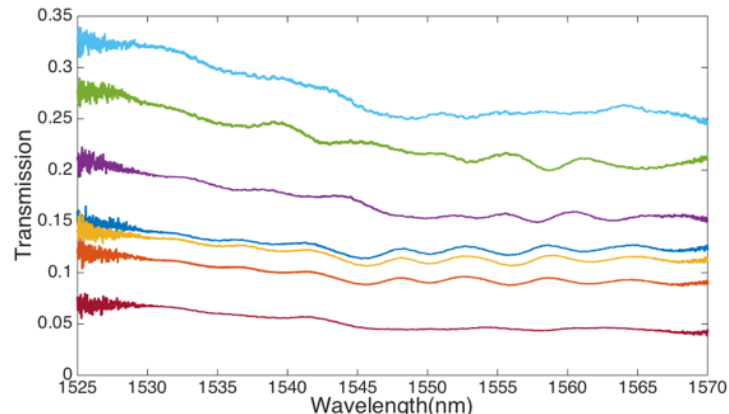

Fig. 2 The transmission result of bend-induced-birefringence Lyot-filter (Different colors refer to different polarization states of polarization controller).

In order to achieve dual-wavelength mode-locked fiber laser, Xin Zhao et al. inserted a programmable attenuator to control the intra-cavity loss to give gain tilt of EDF's gain spectrum. However, because of the difficulty of adjusting the gain tilt, it's difficult to obtain high quality dual-wavelength mode-locking fiber laser by just setting a constant loss. Based on the above studies on the bend-induced-birefringence Lyot-filter, a simpler technique is proposed here for dual-wavelength mode-locked fiber laser generation.

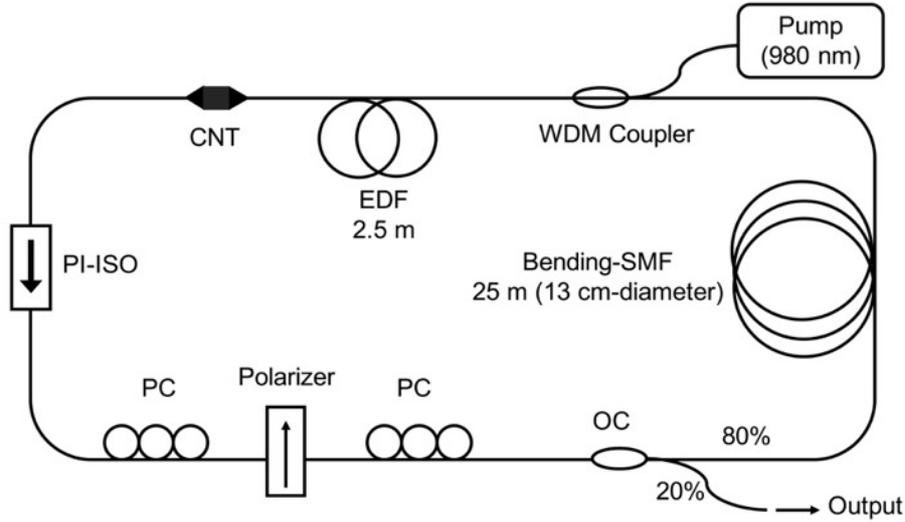

Fig. 3 Schematic diagram of experimental setup (WDM: wavelength-division multiplexer; EDF: erbium-doped fiber; CNT: carbon nanotube; PI-ISO: polarization insensitive isolator; PC: polarization controller; OC: output coupler; SMF: single-mode fiber).

The experimental setup is shown in Fig. 3. The dual-wavelength mode-locked laser is constructed in a ring-cavity configuration, using carbon nanotube (CNT) as the saturable absorber and a bend-induced-birefringence Lyot-filter as a gain tilt spectral equalizer. The gain medium is a segment of 2.5-m-long EDF, which is pumped by a 980 nm laser diode via a 980/1550 wavelength-division multiplexer (WDM) fused coupler. A polarization insensitive isolator (PI-ISO) is spliced in the laser cavity to ensure unidirectional operation. A PC, a polarizer and a section of 25-m-long SMF bent in 13cm diameter coil form a bend-induced-birefringence Lyot-filer. The laser output via the 20% port of a 20/80 fiber coupler while the spectral and temporal characteristics are measured using an OSA and Oscilloscope (Agilent Technologies DS01024A) with a 5GHz Photodiode respectively through 80% port.

## 3. Experimental Details
### 3.1. Single-wavelength mode-locked output

The laser can operate in 3 different regimes, continuous wave (CW) lasing, single-wavelength mode-locking and dual wavelength mode-locking. The output optical spectrum and temporal waveforms of the laser at single-wavelength mode-locking regime at 1532 nm and 1556 nm are shown in Fig. 4 (a)(b) and Fig.4 (c)(d), respectively. The spectral half-width of the laser operating at single-wavelength mode-locking regime at 1532 nm and 1556 nm are 4 nm and 6 nm, respectively. The round trip time in both cases are 150 ns, corresponding to ~7.6MHz, which matches the cavity length of ~30 meters.

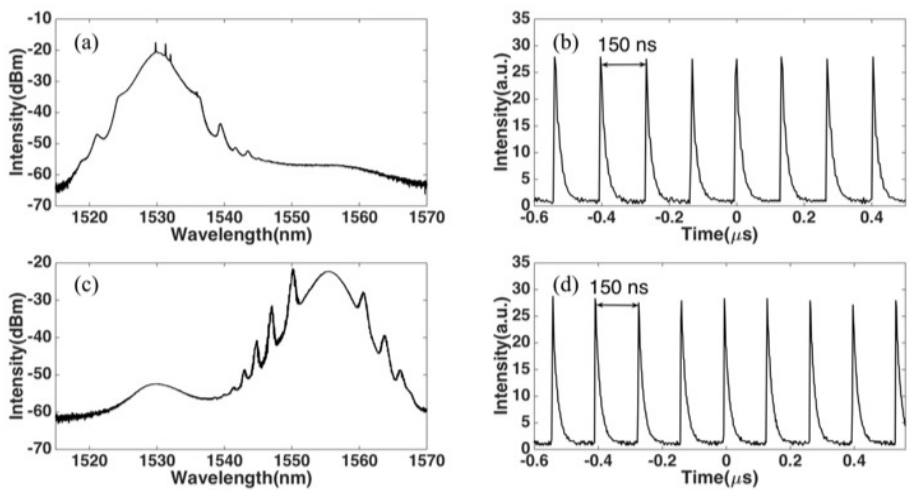

Fig. 4 (a) Optical spectrum (0.01 nm resolution) and (b) temporal waveforms of 1532 nm mode-locked laser (c) Optical spectrum (0.01 nm resolution) and (d) temporal waveforms of 1556 nm mode-locked laser.

### 3.2. Dual-wavelength mode-locked output

By properly adjusting the PCs, dual-wavelength mode-locked output can be obtained. The measured dual-wavelength mode-locked optical spectrum is shown in Fig. 5 (a). The central wavelengths are 1532 nm and 1556 nm and the 3 dB bandwidth are 5 nm and 4.5 nm respectively. CW component can be seen from the spectrum of the pulse at 1532 nm, which maybe attribute to the mode-locking condition is not optimized at 1532 nm. The oscilloscope trace of the dual-wavelength mode-locked laser is shown in the Fig. 5 (b). The repetition time is 150 ns, which is consistent with the cavity length. From the electrical spectrum analyzer trace in Fig. 5 (c), the pulses at 1532 nm have a repetition rate of 7.5981 MHz whilst the the repetition rate of the 1556nm pulses is 7.5986 MHz. Therefore, the repetition rates difference is ~500 Hz, which is due to the group velocity dispersion (GVD) of the laser cavity.

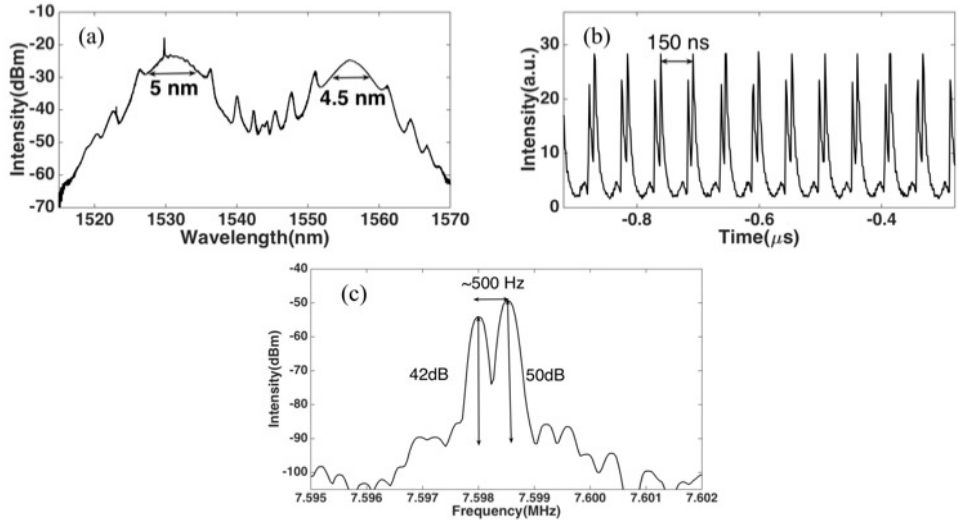

Fig. 5 (a) Optical spectrum (0.01 nm resolution) (b) temporal waveforms and (c) RF spectrum of dual-wavelength mode-locked laser (RBW=40 Hz).

A tunable optical band-pass filter (BPF) (Alnair Labs BVF-200) which has a flat top band-pass and a low in-band chromatic dispersion is used to filter out the one of the two wavelengths, with the pass band set at 1522.3 nm to 1541.2 nm (for 1532nm output) and 1547.1 nm to 1564.7 nm (for 1556 nm output), respectively. The optical spectra and autocorrelation traces of the pulses at 1532 nm and 1556 nm can be measured separately. As shown in the Fig. 6(a) and 6(b), the 1532 nm output pulse has a spectral half-width of ~5 nm and an inferred full-width at half-maximum (FWHM) pulse-width of 940 fs assuming a Gaussian waveform. Because the side wings of the pulse spectrum are filtered out by the band-pass filter, the resulting spectrum resemble more like a Gaussian shape rather than the original hyperbolic-sech shape. Also, as shown in the Fig. 6(c) and (d), the spectral half width and the inferred FWHM pulse-width of the 1556 nm pulse component is 4.5 nm and 820 fs respectively. The time-bandwidth products (TBPs) for the 1532 nm and 1556 nm pulses are 0.504 and 0.457, respectively, which is slightly larger than the transform-limited value of 0.441 assuming Gaussian waveforms. The reason for the non-ideal TBP may be attributed to the the fact that part of the spectral component was filtered out by the BPF. Moreover, because of the high loss induced by the band-pass filter, the measured autocorrelation traces were measured at very low powers impacting the accuracy of the autocorrelation pulse-width measurement.

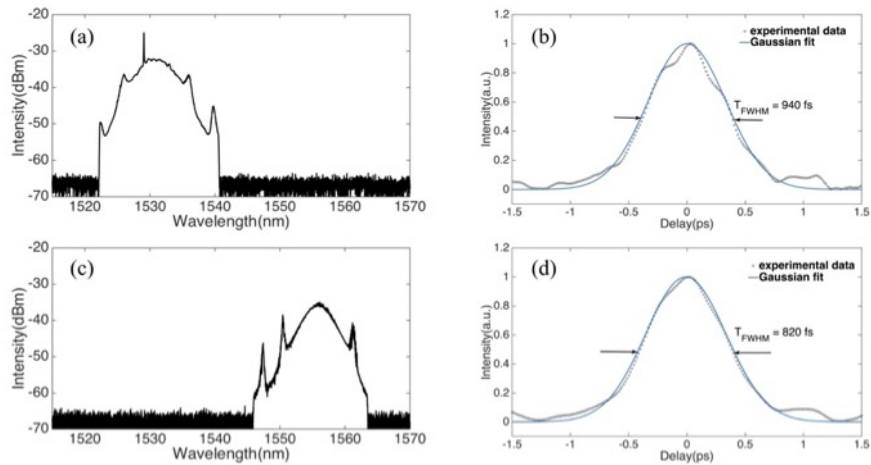

Fig. 6 (a) Optical spectrum (0.01 nm resolution) and (b) autocorrelation trace of the laser output centered at 1532 nm (c) Optical spectrum (0.01 nm resolution) and (d) autocorrelation trace of the laser output centered at 1556 nm.

In order to confirm the functionality of the bend-induced-birefringence Lyot-filter, the polarizer and one of the PCs is intentionally removed from the fiber cavity. It was found that the laser could only operate at a single-wavelength mode-locking regime at 1556 nm despite all effort to adjust the PC. This indicated that without the bend-induced-birefringence Lyot-filter giving the gain tilt to balance the two peaks of the EDF, only a single wavelength will dominate in the pulse formation.

## 4. Conclusions

In this paper, we demonstrate a dual-wavelength mode-locked fiber laser operating at 1532 nm and 1556 nm. The laser employs a CNT saturable absorber and a bend-induced-birefringence Lyot-filer as gain-tilt spectral equalizer to balance the gain competition between the two wavelengths. In addition, the bend-induced-birefringence Lyot-filter can be also used in the laser cavity with other saturable absorbers to achieve single-wavelength mode-locking and dual-wavelength mode-locking operation. This laser may have potential to be used in applications such as dual-comb spectroscopy and Terahertz wave generation.

## Acknowledgements

We are grateful for the support by JSPS Grant-in-Aid for Scientific Research (S) Grant Number 18H05238.